\documentclass[aps,twocolumn,showpacs]{revtex4}
\usepackage{amssymb}
\usepackage{amsfonts}
\usepackage{amsmath}
\usepackage{mathrsfs}
\usepackage{graphicx}

\begin{document}

\title{Energy-momentum tensor is nonsymmetric for spin-polarized photons}

\author{Xiang-Bai Chen$^{a,b}$\footnote{xchen@kku.ac.kr}}

\author{Xiang-Song Chen$^{a}$\footnote{for correspondence: cxs@hust.edu.cn}}

\affiliation{$^a$School of Physics and
MOE Key Laboratory of Fundamental Quantities Measurement,
Huazhong University of Science and Technology, Wuhan 430074, China\\
$^b$Department of Applied Physics, Konkuk University,
Chungju 380-701, Korea}

\date{\today}

\begin{abstract}
It has been assumed for a century that the energy-momentum tensor
of the photon takes a symmetric form,
with the renowned Poynting vector assigned as the same density for
momentum and energy flow.
Here we show that the symmetry of the photon energy-momentum
tensor can actually be inferred from the known difference between the
diffraction patterns of light with spin and orbital angular momentum,
respectively.
The conclusion is that the symmetric expression of energy-momentum tensor
is denied, and the nonsymmetric canonical expression is favored.

\pacs{11.10.Cd, 14.70.Bh, 41.75.Fr}
\end{abstract}
\maketitle

Energy, momentum, and angular momentum are among the most fundamental
quantities in physics. It is awkward that these quantities
can often arouse controversy and confusion.
In hadron physics, for example,
there is no universally accepted scheme to analyze the quark-gluon
origin of the nucleon momentum and spin \cite{Chen08,Chen09,Chen12}.
Even for the familiar photon (or electromagnetic field),
the expression of momentum $\vec P$ and angular
momentum $\vec J$ is problematic under close look.
One often sees the mixed use of two expressions:
\begin{subequations}
\label{J}
\begin{eqnarray}
\vec J&=&\int d^3 x \vec x\times (\vec E\times\vec B)
\label{EB}\\
&=&\int d^3 x \vec x\times E^i\vec\nabla A^i  +\int d^3 x \vec E\times \vec A
\equiv \vec L +\vec S .
\label{SL}
\end{eqnarray}
\end{subequations}
Eq. (\ref{EB}) contains the renowned Poynting vector $\vec E\times\vec B$,
which is derived as the electromagnetic momentum density in common textbooks.
It is however Eq. (\ref{SL}) that separates the intrinsic spin
$\vec S$ from the extrinsic orbital angular momentum $\vec L$.
For a free field, it can be easily shown that Eqs. (\ref{EB}) and (\ref{SL}) give the
same conserved total $\vec J$, and therefore are often regarded as being identical.
However, the momentum and angular momentum densities are after all different in the
two expressions. Our aim is to show that one expression must be wrong at the
density level, and can actually be inferred from the known experiments 
(see Fig. 1 and our explanations below).

\begin{figure}
\includegraphics[width=0.48\textwidth]{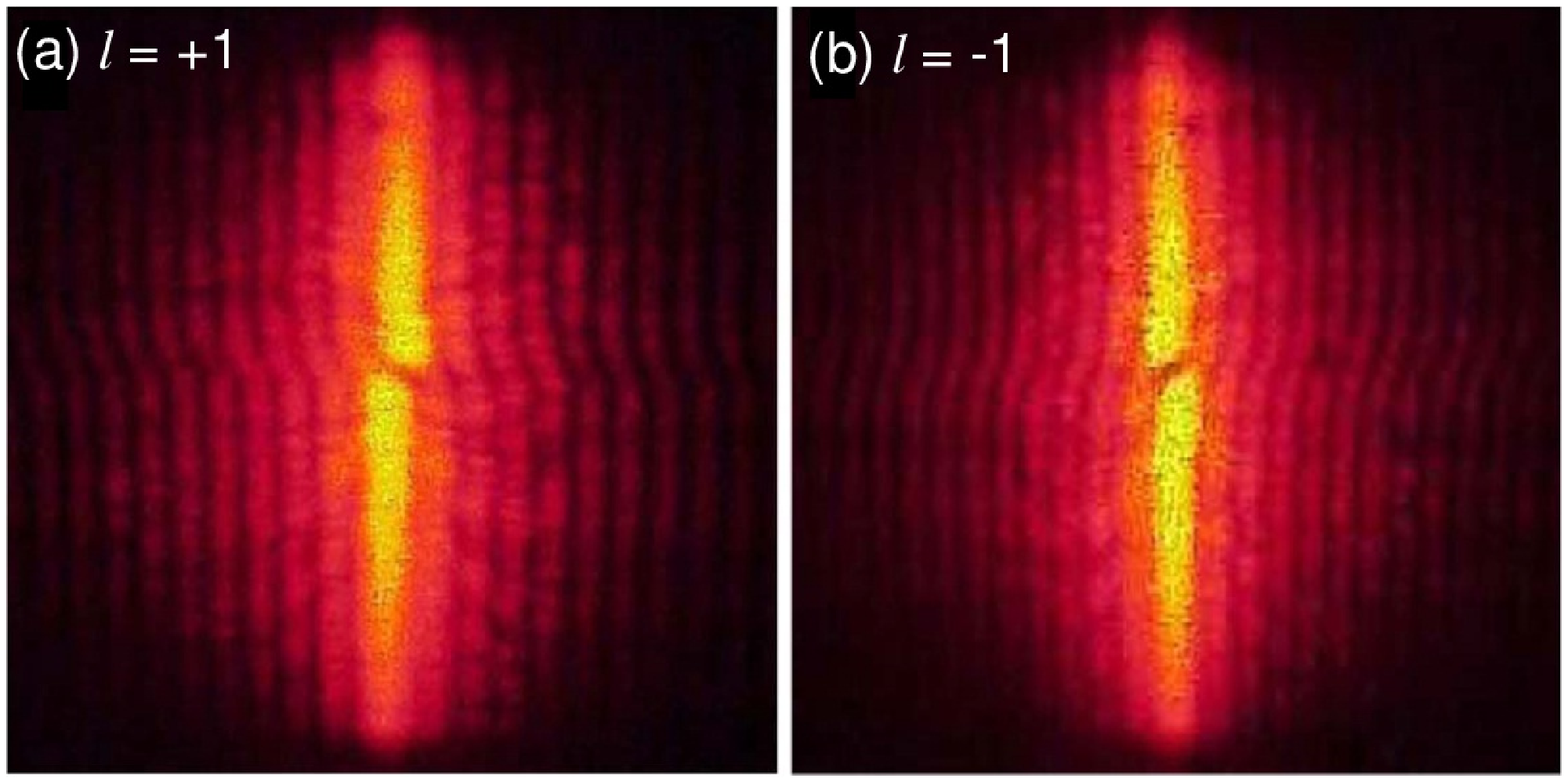}
\includegraphics[width=0.48\textwidth]{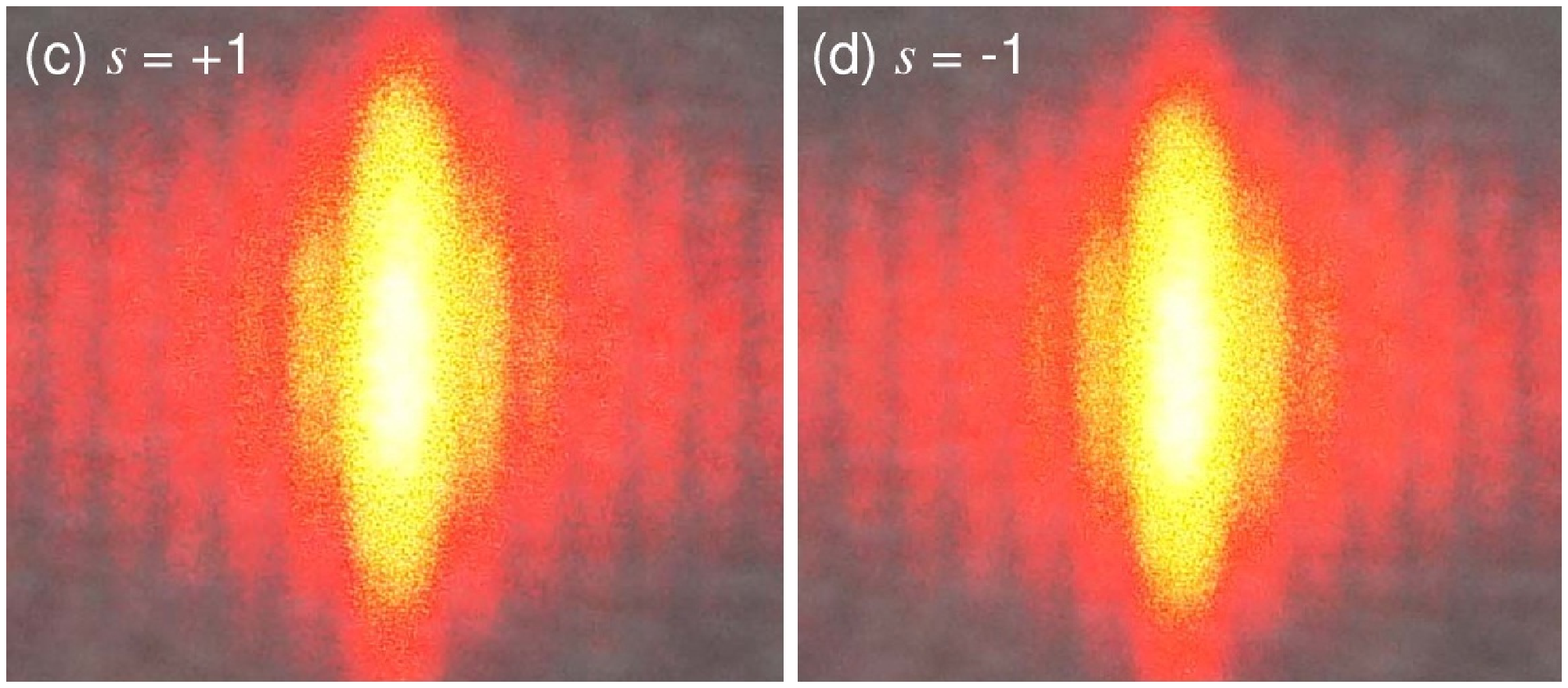}
\caption{\label{measure} Single-slit diffraction patterns
of light with spin $s$ and orbital angular momentum $l$.
(a) $l=+1$, (b) $l=-1$, (c) $s=+1$, (d) $s=-1$. (a) and (b)
are quoted from Fig. 5 of Ref. \cite{Ghai09}. Nonzero
$l$ leads to distortion of the diffraction fringes, while
nonzero $s$ does not. By a careful analysis, this simple 
but critical difference can tell that the 
energy-momentum tensor cannot be symmetric for spin-polarized
photons.}
\end{figure}

Eqs. (\ref{EB}) and (\ref{SL}) correspond to
different expressions of the angular momentum tensor:
\begin{subequations}
\label{M}
\begin{eqnarray}
{\mathscr M}^{\lambda\mu\nu}&=&x^\mu \Theta^{\lambda\nu}-x^\nu \Theta^{\lambda\mu},
\label{scrM}\\
{\cal M}^{\lambda\mu\nu}&=&x^\mu {\cal T}^{\lambda\nu} -x^\nu{\cal T}^{\lambda\mu}
+F^{\lambda\nu}A^\mu-F^{\lambda\mu}A^\nu.
\label{calM}
\end{eqnarray}
\end{subequations}
Here Eq. (\ref{scrM}) gives the total angular momentum with an orbital-like expression,
by using the symmetric energy-momentum tensor
\begin{equation}
\Theta ^{\mu\nu} =F^{\mu\rho} F_\rho^{~\nu} +\frac 14 g^{\mu\nu} F^2
=\Theta ^{\nu\mu},
\end{equation}
while in Eq. (\ref{calM}) the explicit $x$-dependent part is only the
orbital contribution, constructed with the nonsymmetric canonical
energy-momentum tensor
\begin{equation}
{\cal T} ^{\mu\nu} =-F^{\mu\rho} \partial^\nu A _\rho +\frac 14 g^{\mu\nu} F^2
\neq {\cal T} ^{\nu\mu}.
\end{equation}
The two angular momentum tensors are both conserved:
\begin{equation}
\partial_\lambda {\mathscr M}^{\lambda\mu\nu} =\partial_\lambda {\cal M}^{\lambda\mu\nu} =0,
\end{equation}
and give the same angular momentum in Eqs. (\ref{J}):
\begin{equation}
J^k=\frac 12 \epsilon_{ijk} \int d^3x {\mathscr M}^{0ij}=
\frac 12 \epsilon_{ijk} \int d^3x{\cal M}^{0ij}.
\end{equation}

Similarly, the two different energy-momentum tensors are both conserved
and give the same 4-momentum:
\begin{eqnarray}
&\partial_\mu \Theta ^{\mu\nu}=\partial_\mu {\cal T} ^{\mu\nu}=0,\\
&P^\nu =\int d^3x \Theta ^{0\nu} =\int d^3x {\cal T}^{0\nu}. \label{P}
\end{eqnarray}

We will explain the concrete experimental evidence that the
elegant expressions $\Theta^{\mu\nu}$ and ${\mathscr M}^{\lambda\mu\nu}$,
despite their popularity, are really wrong.
In Ref. \cite{Chen12a}, we already gave a hint by examining the
energy-flow component that for polarized
electrons the nonsymmetric canonical energy-momentum tensor is favored
over the symmetric one, and proposed an experimental test.
The energy-flow component is nevertheless unable to discriminate
$\Theta ^{\mu\nu}$ from ${\cal T}^{\mu\nu}$ for the photon \cite{Chen12a}.
In this paper, we look at the more delicate momentum-flow component.

Consider a light beam propagating along the $z$ axis, around which the beam
is rotationally symmetric. 
Suppose that $T^{\mu\nu}$ is the true energy-momentum tensor of the light beam.
The components $T^{zx}$ and $T^{zy}$ describe the flow of $P^x$ and $P^y$
along the $z$ direction. An often ignored but vital fact is that
$T^{zx}$ and $T^{zy}$ are measurable locally. Especially, we will see that for
the sake of telling the symmetry of $T^{\mu\nu}$, it suffices to make a
rough estimation of the momentum flux by looking at the
diffraction fringes after passing through a small aperture.
Hence, the energy-momentum tensor does not have the usually assumed
arbitrariness. This warns us that $\Theta ^{\mu\nu}$ and ${\cal T}^{\mu\nu}$
cannot be both correct, and can in principle be discriminated by comparing
the measured density of momentum flow with that calculated via $\Theta ^{\mu\nu}$ 
and ${\cal T}^{\mu\nu}$. In what follows, we present a clever way to experimentally
distinguish $\Theta ^{\mu\nu}$ from ${\cal T}^{\mu\nu}$, with no need to know the
detailed profile or wavefunction of the beam.

The technique is to look at a very elucidating quantity
\begin{equation}
K^z_M=\int M^{zxy}dxdy=\int (xT^{zy}-yT^{zx})dxdy. \label{Kz}
\end{equation}
Here the integration is over the beam cross section, say, in the $x$-$y$ plane.
If $T^{\mu\nu}=\Theta^{\mu\nu}$, then $K^z_M$ measures the flow of the
{\em total} angular momentum $J^z$ across the $x$-$y$ plane.
But if $T^{\mu\nu}={\cal T}^{\mu\nu}$, then $K^z_M$ measures only
the flow of the {\em orbital} angular momentum $L^z$ across the $x$-$y$ plane.

By Eq. ({\ref{Kz}), the angular-momentum flow $K^z_M$ is the {\em moment}
of the momentum flow $T^{iz}$. 
To get a nonzero $K^z_M$ by the integration in Eq. (\ref{Kz}), the momentum
flow $T^{zx}$ and $T^{zy}$ must display a circular behavior around the $z$ axis.
For example, to give a positive $K^z_M$, the dominating configuration must be  
$T^{zy}>0$ for $x>0$ and $T^{zy}<0$ for $x<0$ 
(and similarly, $T^{zx}<0$ for $y>0$ and $T^{zx}>0$ for $y<0$). 
In this case, therefore, (remembering the physical meaning of momentum flow,) 
the photons passing 
though the region with $x>0$ must have a net positive $P^y$, while those through
$x<0$ must have a net negative $P^y$. 
(Similarly, the diffracted photons through $y>0$ must have a net negative $P^x$, 
while those through $y<0$ must have a net positive $P^x$.)
Such net momentum would manifest in diffraction pattern of light:
the diffraction fringe would shift towards $+y$ ($-y$) direction for $x>0$ ($x<0$),
or shift towards  $+x$ ($-x$) direction for $y<0$ ($y>0$), 
leaving a distorted diffraction pattern.

If $T^{\mu\nu}=\Theta^{\mu\nu}$, then the above distortion
would be observed as long as the beam carries a nonzero $J^z$, no matter of
spin and/or orbital origin.
On the other hand, such a distortion would only be observed for a beam with
nonzero $L^z$ (whatever $S^z$ is) if $T^{\mu\nu}={\cal T}^{\mu\nu}$.
Therefore, the patterns of the simple single-slit diffraction of light
with spin and orbital angular momentum can tell concretely
whether $T^{\mu\nu}$ can be symmetric or not.

In fact, these diffraction patters are already {\em known}:
For nonzero $L^z$ one does observe
the expected distortion (see, e.g., Fig. 5 of Ref. \cite{Ghai09}, which
is quoted here in Fig. 1), while circularly polarized light carrying nonzero
$S^z$ but zero $L^z$ does not display any similar distortion (see Fig. 1).
We therefore conclude that for spin-polarized photons the symmetric expression
of energy-momentum tensor is excluded by experiment, and the nonsymmetric
canonical expression is favored.

We close this paper with the following remarks:

(i) The same analysis can be performed for the electron, by studying
the diffraction patterns of spin-polarized beam and the recently
realized electron beam with orbital angular momentum
\cite{McMo11,Verb10,Uchi10}; and the same conclusion can be expected.

(ii) From our illustration, spin-polarization does not produce a 
circular momentum, but it does produce a circular energy flow \cite{Chen12a}.
In the canonical expression, momentum and energy flow are two different 
quantities.

(iii) The canonical expressions are in general gauge-dependent and need
gauge-invariant revision. Such revision is subject to certain theoretical
uncertainties \cite{Chen12}, but experiments would ultimately remove such
uncertainties.

(iv) In the interacting case, the two expressions in Eqs. (1) no longer
give the same angular momentum, and those in Eq. (\ref{P}) do not
give the same momentum, either. For a strongly interacting system the
difference can even be huge \cite{Chen09}.

(v) Given the experimentally selected nonsymmetric canonical energy-momentum
tensor ${\cal T}^{\mu\nu}$, one has to seriously consider whether Nature
would choose to use ${\cal T}^{\mu\nu}$ for gravitational coupling, and thus
abandon the Einstein's theory (which requires a symmetric energy-momentum tensor);
or, Nature would favor Einstein's theory and permits both the symmetric
$\Theta^{\mu\nu}$ and the nonsymmetric ${\cal T}^{\mu\nu}$, probably with the latter 
describing only the inertial energy-momentum tensor.

This work is supported by the China NSF Grant Nos.
11275077 and 11035003, the NCET Program of the China
Ministry of Education, and the Korea NRF Grant No. 2010-0022857.

\end{document}